# Convergence of a Periodic Orbit Family Close to Asteroids During a Continuation

Haokun Kang[1], Yu Jiang[1], Hengnian Li[1]

1. State Key Laboratory of Astronautic Dynamics, Xi'an Satellite Control Center, Xi'an 710043, China

Y. Jiang (✉) e-mail: jiangyu_xian_china@163.com (corresponding author)

H. Li (✉) e-mail: henry_xscc@aliyun.com (corresponding author)

**Abstract**. In this work, we study the continuation of a periodic orbit on a relatively large scale and discover the existence of convergence under certain conditions, which has profound significance in research on asteroids and can provide a total geometric perspective to understanding the evolution of the dynamic characteristics from a global perspective. Based on the polyhedron model, convergence is derived via a series of theoretical analyses and derivations, which shows that a periodic orbit will evolve into a nearly circular orbit with a normal periodic ratio (e.g., 2:1, 3:2, and 4:3) and almost zero torsion under proper circumstances. As an application of the results developed here, three asteroids, (216) Kleopatra, (22) Kalliope and (433) Eros, are studied, and several representative periodic orbit families are detected, with convergence in three different cases: bidirectional, increasing-directional and decreasing-directional continuation. At the same time, four commonalities among these numerical examples are concluded. First, a (pseudo) tangent bifurcation arises at the cuspidal points during the variations in the periodic ratios in a single periodic orbit family. In addition, these cuspidal points in the periodic ratio coincide with the turning points during the variations in the average radius, the maximal torsion and the maximal radius of curvature. Furthermore, the periodic ratio increases (or decreases) with an increase (or decrease) in the Jacobian constant overall. We find the relationship between periodic ratio and Jacobian constant. The results implies that the periodic ratios for a fixed resonant status have an infimum and a supremum. Finally, if the periodic orbit converges to a point, it can be an unstable equilibrium point of the asteroid.

## 1. Introduction

With the rapid development of space exploration technology, asteroids have become a hot area of research for decades due to the increased number of space missions. As one of the important branches, periodic orbits play a significant role and are considered to be key to understanding the nature of asteroid dynamics. Therefore, various research that regards this issue has been obtained, such as gravity potential field computation[1-8], equilibrium points searching and analyses [9-15], generation of periodic orbits [16-20], classification [21] and bifurcation analysis [22-23]. These outstanding studies have further contributed to the development of asteroid research and lay a firm foundation for later studies. From these previous studies and through further research, we discovered convergence in the periodic orbit family during its continuation with a change in the Jacobian constant on a large scale, and we derive the existence of this convergence from a theoretical perspective. This conclusion is also validated by several representative periodic orbit families with different types of convergence in the adjacent area of three different asteroids. At the same time, some hidden characteristics of commonalities are also revealed.

Asteroids are considered to be remnants of materials that were created by the Sun and the planets [24], and almost all of them have irregular shapes, which poses a great challenge when performing modelling to approximate their potential field. Previously, numerical experiments and studies employed dynamics models with similar regular shapes, such as a straight segment [25-26], a solid circular ring [27], a homogeneous annulus disc [28-29], and a homogeneous cube [11]. As observation technologies advance, more data are gathered, which can help realise the establishment of an accurate model. Later on, the polyhedron model for the gravity potential field computation of an asteroid, which was proposed by Werner (1994)[1], came into public view and became of great interest. Using this model, research and analyses on periodic orbits can be conducted in a more rigorous manner.

In their epoch-making work, Yu and Baoyin [17] proposed the hierarchical grid searching method based on a given Jacobian constant with a polyhedron model. To search for periodic orbits, there are several methods proposed, like using the indirect method to obtain the surrounding orbits near an asteroid represented by a simple shape [18]. After that, a number of periodic orbits with accurate descriptions were generated near many asteroids, such as (216) Kleopatra [17], (433) Eros [30], (101955) Bennu [31], (6489) Golevka and Comet 1P/Halley [21]. Although this description for a periodic orbit is direct and can be visualised, the inner properties and dynamic characteristics are still unclear. Jiang et al. [21] utilised the topological distribution of characteristic multipliers for each periodic orbit and classified them into 13 cases based on the symplectic eigenvalue theorem. Simultaneously, the stabilities of these cases can be distinguished.

A periodic orbit can also be generated and continued from a known one. In other words, a periodic orbit can be extended numerically into one mono-parametric set with changing parameters of the Jacobian constant [17]. These periodic orbits compose a periodic orbit family and can reflect the dynamic properties of an asteroid continuously. Therefore, exploration of periodic orbit families attracts a large amount of attention and yields many new research topics.

One of the most important characteristics of a single periodic orbit family is bifurcation. Originally, Yu and Baoyin [17] discussed the variations in Floquet multipliers in a periodic orbit family near asteroid (216) Kleopatra, which revealed the changes in the stability directly. After Jiang et al. [21] completed the topological structure classification of periodic orbits, four basic types of bifurcations, namely, tangent bifurcations, period-doubling bifurcations, Neimark-Sacker bifurcations, and real saddle bifurcations, were summarised, making this analysis step a highly useful stage. Furthermore, multiple bifurcations were also discovered in the continuation of the periodic orbit families near (216) Kleopatra [22] and (433) Eros [23]. The corresponding pseudo bifurcations were also discussed with regard to (22) Kalliope.

Another critical feature of a single periodic orbit family is the periodic ratio, which is defined as the proportion of the orbital periods and the asteroid's rotation period. Yu and Baoyin [32] found that there were many resonant orbits distributed in the near-field of asteroid (216) Kleopatra, which meant that these periodic orbits possessed normal periodic ratios (e.g., 2:1, 3:2, and 4:3). Later, several periodic orbits with different normal periodic ratios near asteroid (101955) Bennu and (216) Kleopatra were discovered [31]. This finding showed that the periodic ratio was not strictly nominal but fell into an interval around the nominal ratios, which means that the periodic ratio might change slightly around the normal periodic ratio with the variation in the Jacobian constant.

In fact, all of the aforementioned results on bifurcation and the periodic ratio are short-term variations of the Jacobian constant, which can be called the local characteristics. However, when considering long-term variations of the Jacobian constant and global properties for a single periodic orbit family, a peculiar phenomenon is discovered. In our paper, we conduct the continuation of a long-term variation and focus on the global properties of a single periodic orbit family; then, we find that periodic orbits in a single family can present a type of convergence on certain conditions when the Jacobian constant is either increasing or decreasing.

We derive and prove the existence of this convergence theoretically based on the model of a polyhedron, whereby a corollary is drawn. Specifically speaking, under proper conditions, the periodic orbit will evolve into nearly circle orbits with normal periodic ratios. Moreover, we also utilise the radius of curvature and the torsion to quantise the characteristics of a periodic orbit, which provides a geometric view to show this long-term variation in a single periodic orbit family. Afterward, to make the finding convincing, this corollary is applied and verified in three different asteroids, (216) Kleopatra, (22) Kalliope, and (433) Eros. The shapes of these three asteroids are dumbbell-shaped, Potato-shaped, and bow-shaped, which are representative. Furthermore, several commonalities among these numerical examples from different asteroids are concluded. First, the periodic ratio will present sharp changes during the continuation of the Jacobian constant, where the cuspidal points form and (pseudo) tangent bifurcation occurs. In addition, these cuspidal points in the periodic ratio coincide with the turning points of the average radius, the maximal torsion and the maximal radius of curvature. Furthermore, the overall trend of the periodic ratio agrees with one of the Jacobian constants. This finding means that the periodic ratio increases (or decreases) with an increase (or decrease) in the Jacobian constant on the whole. Finally, if a periodic orbit shrinks to a point during the continuation, it can be an unstable equilibrium point of the corresponding asteroid [33].

To illustrate our research more convincingly and explicitly, we structure this paper as follows. In Section 2, basic knowledge about the gravitational potential, dynamic equations, characteristic multipliers and concepts of curvature and torsion are introduced. Section 3 presents the related classifications of periodic orbits and four basic (pseudo) bifurcations. Section 4 derives and verifies the reasonability of the convergence in a single periodic orbit family under certain conditions theoretically, and a corollary is summarised as a conclusion. Then, this corollary is applied in Section 5 to verify the existence of convergence in many periodic orbit families near three different asteroids. At the same time, the commonalities among these numerical examples are analysed. Finally, the conclusions are drawn in Section 6.

## 2. Basic dynamic characteristics

In this section, the gravitational potential, motion equations and characteristic multipliers of periodic orbits will be discussed and recalled as the basic dynamic characteristics for the whole study.

## 2.1 Gravitational potential

Usually, the gravity field of an asteroid presents strong irregularity near its surface due to its irregular shape. To obtain a more approximate dynamic simulation, the model of a polyhedron with constant density for an asteroid is adopted in this study. Then, the gravitational potential $U$ can be expressed as the summation of the finite terms associated with the faces and edges [34]:

$$U(\mathbf{r}) = \frac{1}{2} G\sigma \sum_{e \in edges} (\mathbf{r}_e \cdot \mathbf{E}_e \cdot \mathbf{r}_e) \cdot L_e - \frac{1}{2} G\sigma \sum_{f \in faces} (\mathbf{f}_e \cdot \mathbf{F}_e \cdot \mathbf{f}_e) \cdot \omega_f, \quad (1)$$

and its gradient can be derived as

$$\nabla \mathbf{U} = -G\sigma \sum_{e \in edges} (\mathbf{r}_e \cdot \mathbf{E}_e) \cdot L_e + G\sigma \sum_{f \in faces} (\mathbf{f}_e \cdot \mathbf{F}_e) \cdot \omega_f. \quad (2)$$

Here, $G$ denotes the gravitational constant, and $\sigma$ indicates the density of the asteroid. The body-fixed vectors from the field point at $\mathbf{r}$ to arbitrary points on the edge $e$ and face $f$ are labelled $\mathbf{r}_e$ and $\mathbf{f}_e$. $\mathbf{E}_e$ and $\mathbf{F}_e$ present geometric parameters of the edges and faces, respectively. $L_e$ is defined by the integration factor of the particle position and edge $e$, and it has the form of

$$L_e = \int_e \frac{1}{r} ds = \ln \frac{a+b+l_e}{a+b-l_e}, \quad (3)$$

where $a$ and $b$ are the distance between the field point $\mathbf{r}$ and two ends of the edge $e$ and $l_e$ is the length of edge $e$. Here, $\omega_f$ stands for the solid angle of the face $f$ relative to the particle, which is defined as

$$\omega_f = \iint_S \frac{\Delta z}{r^3} dS, \quad (4)$$

Here, $\Delta z$ presents the distance from the field point to the face $f$.

It is clear that $L_e$ and $\omega_f$ are fixed values for arbitrary edge $e$ and face $f$ once the field point is selected.

## 2.2 Dynamic equations

For an asteroid with rotational angular velocity $\mathbf{\Omega}$ relative to the inertial space, with point $\mathbf{r}$ being the radius vector from the asteroid's centre of mass to the particle, the equation of motion can be expressed as a second-order ordinary differential equation [35]:

$$\ddot{\mathbf{r}} + 2\mathbf{\Omega} \times \dot{\mathbf{r}} + \mathbf{\Omega} \times (\mathbf{\Omega} \times \mathbf{r}) + \dot{\mathbf{\Omega}} \times \mathbf{r} + \frac{\partial U(\mathbf{r})}{\partial \mathbf{r}} \tag{5}$$

where $U(\mathbf{r})$ is the gravitational potential in Eq. (1). The body-fixed frame [36] can be defined by an orthogonal right-handed set of unit vectors $\{\mathbf{e}\} = \{\mathbf{e}_x, \mathbf{e}_y, \mathbf{e}_z\}^T$, and Eq. (5) can be written in scalar form:

$$\begin{cases} \ddot{x} + \dot{\Omega}_y z - \dot{\Omega}_z y + 2\Omega_y \dot{z} - 2\Omega_z \dot{y} - \Omega_y^2 x - \Omega_z^2 x + \Omega_x \Omega_y y + \Omega_z \Omega_x z + \frac{\partial U}{\partial x} = 0, \\ \ddot{y} + \dot{\Omega}_z x - \dot{\Omega}_x z + 2\Omega_z \dot{x} - 2\Omega_x \dot{z} - \Omega_x^2 y - \Omega_z^2 y + \Omega_y \Omega_z z + \Omega_x \Omega_y x + \frac{\partial U}{\partial y} = 0, \\ \ddot{z} + \dot{\Omega}_x y - \dot{\Omega}_y x + 2\Omega_x \dot{y} - 2\Omega_y \dot{x} - \Omega_x^2 z - \Omega_y^2 z + \Omega_x \Omega_z x + \Omega_y \Omega_z y + \frac{\partial U}{\partial z} = 0. \end{cases} \tag{6}$$

If the unit vector $\mathbf{e}_z$ is defined as $\mathbf{e}_z = \mathbf{\Omega}/\Omega (\Omega = \|\mathbf{\Omega}\|)$, for the uniformly rotating asteroid, the dynamical equations in Eq. (6) can be reorganised as

$$\begin{cases} \ddot{x} - 2\Omega \dot{y} - \Omega^2 x + \frac{\partial U}{\partial x} = 0, \\ \ddot{y} - 2\Omega \dot{x} - \Omega^2 y + \frac{\partial U}{\partial y} = 0, \\ \ddot{z} + \frac{\partial U}{\partial z} = 0. \end{cases} \tag{7}$$

After introducing the state vector $\mathbf{x} = \{x, y, z, \dot{x}, \dot{y}, \dot{z}\}^T$, it is easy to transform Eq. (7) into a first-order ordinary differential with the form of

$$\mathbf{x} = \mathbf{f}(\mathbf{x}). \tag{8}$$

The solutions to the equation $\mathbf{f}(\mathbf{x}) = 0$ are called the equilibrium points of the asteroid.

According to the eigenvalues of the coefficient matrix after the local linearisation of Eq. (7), Jiang et al. (2014) [33] classified the equilibrium points into five cases, whereby one of them is linear stable and four of them are unstable. Table 1 lists these five cases, where the names and abbreviations are inherited from ref. [13].

Table 1 The topological manifold classification of equilibrium points.

| Case | Eigenvalues | Stability |
| --- | --- | --- |
| Case 1 | $\pm i\beta_j (\beta_j \in R^+; j=1,2,3)$ | LS |
| Case 2 | $\pm \alpha_j (\alpha_j \in R^+; j=1), \pm i\beta_j (\beta_j \in R^+; j=1,2)$ | U |
| Case 3 | $\pm \alpha_j (\alpha_j \in R^+; j=1,2), \pm i\beta_j (\beta_j \in R^+; j=1)$ | U |
| Case 4a | $\pm \alpha_j (\alpha_j \in R^+; j=1), \pm \sigma \pm i\tau (\sigma, \tau \in R^+)$ | U |
| Case 4b | $\pm \alpha_j (\alpha_j \in R^+; j=1,2,3)$ | U |
| Case 5 | $\pm \sigma \pm i\tau (\sigma, \tau \in R^+), \pm i\beta_j (\beta_j \in R^+; j=1)$ | U |

Note: LS stands for linear stable, U stands for unstable.

**2.3 Characteristic multipliers**

Denote $S_P(T)$ as the set of periodic orbits with period $T$. For any periodic orbit $\forall p \in S_p(T)$, combined with Eq. (7), the state transition matrix can be defined as

$$\phi(t) = \int_0^t \frac{\partial \mathbf{f}}{\partial \mathbf{x}}[p(\tau)]d\tau. \qquad (9)$$

Then, the monodromy matrix with the form of

$$\mathbf{M} = \phi(T) \qquad (10)$$

presents the transition of the state along the whole periodic orbit $p \in S_p(T)$. The eigenvalues of this matrix are the characteristic multipliers of the periodic orbit $p$; they are called **Floquet multipliers** and are fixed to the specified periodic orbit $p$. For this asteroid dynamic system, $\mathbf{M}$ has been proven to be symplectic, which means that if $\lambda$ is one Floquet multiplier of $\mathbf{M}$, then $\lambda^{-1}$, $\bar{\lambda}$, and $\bar{\lambda}^{-1}$ are also Floquet multipliers of $\mathbf{M}$. Considering that $p$ is a periodic orbit, we can conclude that $+1$ is a Floquet multiplier of $M$ and has a multiplicity of at least 2.

## 2.4 The radius of curvature and torsion

Considering that the periodic orbits surrounding the asteroid are differential geometry curves in three dimensions, we can introduce the curvature as

$$\boldsymbol{\kappa} = \frac{\dot{\mathbf{r}} \times \ddot{\mathbf{r}}}{|\dot{\mathbf{r}}|^3}. \tag{11}$$

Hence, the radius of curvature can be expressed as

$$\rho = \frac{1}{|\boldsymbol{\kappa}|}. \tag{12}$$

The torsion of a curve, which measures how sharply an orbit is twisting out of the plane of curvature, can be defined as

$$\tau = \frac{\det(\dot{\mathbf{r}}, \ddot{\mathbf{r}}, \dddot{\mathbf{r}})}{\|\dot{\mathbf{r}} \times \ddot{\mathbf{r}}\|^2}. \tag{13}$$

These quantities can represent the characteristics of a periodic orbit from a geometric aspect, and they provide a brand-new analysis perspective.

## 2.5 Classifications and numerical continuation of periodic orbits

The periodic orbit is one of the important characteristics during the study of the dynamic motion near an asteroid. Yu and Baoyin [17] proposed the hierarchical grid searching method to seek periodic orbits. Many periodic orbits are generated near different asteroids. In this section, we will introduce some basic classifications of periodic orbits and the basic (pseudo) bifurcations of the periodic orbit families during numerical continuation.

Based on the conception of Floquet multipliers and their distribution on a complex plane, Jiang et al. [21] classified the periodic orbits into 13 different cases in the potential field of a rotating celestial body. To facilitate comparison, six cases with abbreviations that will be mentioned in our paper are plotted in Fig. 1.

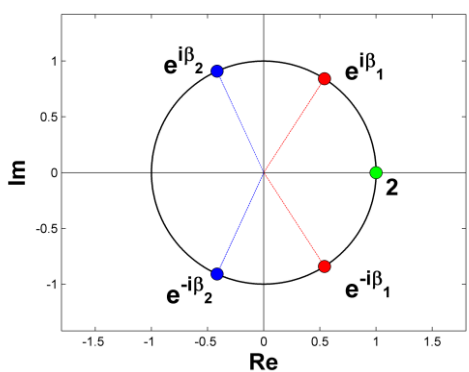

(a) Case P2

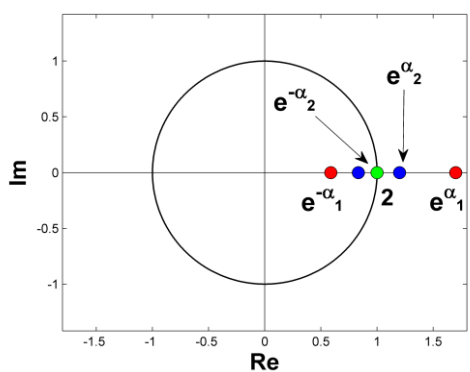

(b) Case P3

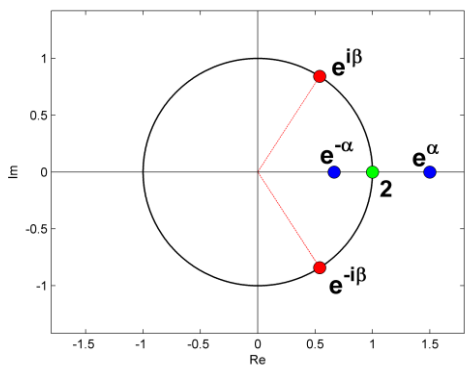

(b) Case P4

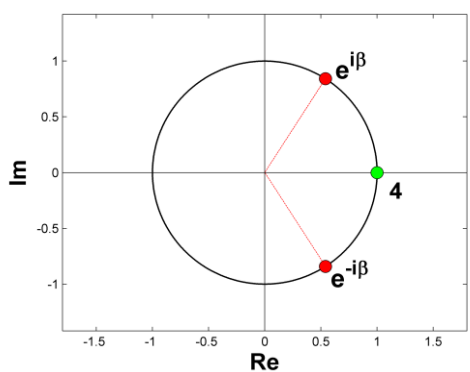

(d) Case P5

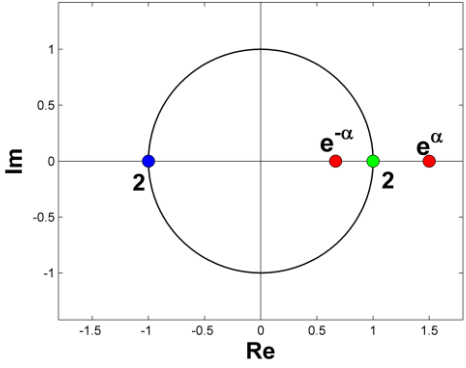 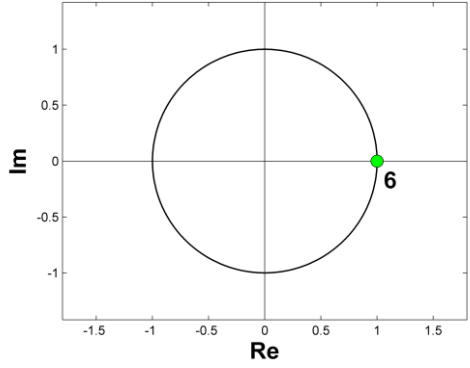

(e) Case P6    (f) Case P7

**Fig. 1**  Six related cases of classification.

By choosing the Jacobian constant as the continuation parameter, one periodic orbit can be extended numerically into one mono-parametric family of periodic orbits [17]. During this procedure, the topological structure can change when the Floquet multipliers collide, which brings about bifurcations. On the other hand, if the topological structure does not change when the Floquet multipliers coalesce, pseudo bifurcations occur. Jiang et al. [21] found that there were four basic (pseudo) bifurcations, namely, (pseudo) period-doubling bifurcations, (pseudo) tangent bifurcations, (pseudo) real saddle bifurcations, and (pseudo) Neimark-Sacker bifurcations, for an irregular body. Accordingly, when the continuation procedure is sufficiently elongated, multiple bifurcations will arise. For instance, triple bifurcations consisting of one period-doubling bifurcation and two real saddle bifurcation were found during the continuation of a periodic orbit family around asteroid (433) Eros [23].

## 3 Convergence of the periodic orbit family

During the numerical continuation in one periodic orbit family, the Jacobian constant can fall into a local valley and become stuck, or it could change steadily. The former case can stop the continuation, but the latter case will prompt the periodic orbit to evolve into a nearly circular state, which presents a type of convergence. In this section, we will focus on the latter case and prove this convergence theoretically.

The Jacobian constant stands for the total energy of one periodic orbit. Combining the numerical continuation process in Yu and Baoyin [17], it is intuitive to find that the average radius of the orbit could increase due to the variety of the Jacobian constant. Once the orbit is expanded sufficiently far from the asteroid, the gravitational potential based on the model of a polyhedron can degenerate into a point source field, which will fail to present the irregularity of the gravitational potential. Let us conduct this discussion more theoretically.

When the distance between the field point and the asteroid is much larger than the maximum diameter of the asteroid during the continuation, $a$ and $b$ in Eq. (3) will be much larger than $l_e$ for any edge $e$. This circumstance can lead to having item $a+b+l_e$ be almost equal to $a+b-l_e$, and the $\Delta z$ in Eq. (4) is nearly zero. Then, it is easy to determine that $L_e \approx 0$ and $\omega_f \approx 0$. Combining the explicit expression of the gravitational potential in Eq. (1) and its gradient in Eq. (2), we can conclude that

$$U \approx 0, \quad \nabla U \approx 0 \tag{14}$$

By multiplying the first equation in Eq. (7) with $\dot{x}$, adding the second equation multiplied by $\dot{y}$ and the third equation multiplied with $\dot{z}$, and then integrating once, we obtain the following:

$$\dot{x}^2 + \dot{y}^2 + \dot{z}^2 - \omega^2(x^2 + y^2) + U = C_1, \tag{15}$$

Here, $C_1$ is an integral constant.

When a particle is far from the asteroid, the gravitational potential and the corresponding gradient at this field point admit Eq. (14). Combining Eqs. (7) and (15), it follows that

$$\begin{cases} \dot{z} = C_2 \\ \dot{x}^2 + \dot{y}^2 - \omega^2(x^2 + y^2) \approx C_1 - C_2 \end{cases} \quad (16)$$

Here, $C_2$ is an integral constant. Considering the symmetry of the gravitational potential at the field point far from the asteroid, it follows that

$$C_2 = 0, \quad z = 0. \quad (17)$$

Substituting into Eq. (16), it is obvious to find that the circular orbit is one of the solutions, which means that the final state of the periodic orbit during the continuation can be a circular orbit in the equatorial plane.

For further analysis, we assume that the velocity of a particle along this circular orbit is $v$ and the average radius is $R$, and we substitute them into Eq. (16); then, we have

$$v^2 \approx \omega^2(R^2 + C_3), \quad (18)$$

Here, $C_3 = (C_1 - C_2)/\omega^2$. Suppose that this periodic orbit goes $N$ times around the asteroid; then, its period $T_p$ can be given as

$$T_p = \frac{2\pi R N}{v}. \quad (19)$$

Combining Eq. (18), we can obtain

$$\frac{T_p}{T} \approx N\sqrt{\frac{R^2}{R^2 + C_3}}, \quad (20)$$

where $T = 2\pi/\omega$ is the period of the asteroid. Therefore, it is not difficult to conclude that the periodic ratio between the particle and the asteroid will converge to the integer $N$ as the average radius expands, i.e. $\lim\limits_{R \to \infty} \frac{T_p}{T} = N$.

Therefore, combining Eqs. (17) and (20), we can conclude the following corollaries for the convergence of the periodic orbit during the continuation:

**Corollary 1** *If the continuation of a periodic orbit based on the Jacobian constant could always be conducted, then a periodic orbit will converge to a nearly circular periodic orbit in the equatorial plane with the multiplicity of an integer, and the periodic ratio will converge to that integer.*

Considering the final state of the periodic orbit and its geometric property in Section 2.4, it should be inferred that the maximal radius of curvature will increase steady, and the maximal torsion will decrease to zero finally, which will be validated by several periodic orbit families in the next section.

**4 Different directions for continuations in applications**

After many numerical calculations, we determined that the convergence described in Corollary 1 exists in many periodic orbits near asteroid (216) Kleopatra, asteroid (22) Kalliope and asteroid (433), whose physical parameters are period (P), diameter (D) and density (ρ) [37-41], which are listed in Table 2. It is noteworthy that some of these periodic orbits present this type of convergence with a change in the Jacobian constant from two different directions, increasing and decreasing. However, for some other periodic orbits, this convergence can only be discovered when the continuation is either increasing or decreasing. This means that the Jacobian constant will stop at a local extremum or the periodic orbits will gradually vanish into a point if the Jacobian constant changes along the other direction. Therefore, the main discussion in this section will be conducted around these three different cases of continuation, bidirectional continuation, increasing-directional continuation and decreasing-directional continuation.

**Table 2** Physical parameters for three asteroids.

| Asteroids | P(h) | D(km) | ρ(g·cm$^{-3}$) |
|---|---|---|---|

| (216) Kleopatra | 5.385 | 219.04 | 3.6 |
| (22) Kalliope | 4.148 | 191.94 | 3.35 |
| (433) Eros | 5.27 | 34.4 | 2.67 |

## 4.1 Bidirectional continuation

In this case, we mainly consider the periodic orbit families that present the phenomenon of convergence in Corollary 1 with the changes in the Jacobian constant along two directions, increasing and decreasing.

### 4.1.1 (216) Kleopatra

The periodic orbit with an initial position of [-0.7496, -0.4121, 0.1982][1] and velocity of [-3.4200, 4.6317, -2.4383] is very representative near asteroid (216) Kleopatra. To perform further analyses, we calculate the periodic ratio, the average radius, the maximal radius of curvature and the maximal torsion of each periodic orbit, and we plot their variations versus the continuation from two directions in Fig. 2.

Figure 2(a) plots the change tendency of the periodic ratio versus the Jacobian constant, which can show that the periodic ratio will diminish from 2:1 to 1:1 slowly with a decrease in the Jacobian constant and rise from 2:1 to 3:1 gradually with an increase in the Jacobian constant. Specifically, when the Jacobian constant decreases to the value of $-2.2851\times10^3 \cdot m^2 \cdot s^{-2}$, a cuspidal point turns up, and the periodic ratio is 1.9129:1 at that time, which is marked as Cuspidal Point 1. Afterward, the periodic ratio deceases rapidly and converges to 1:1 gradually with a fading velocity.

---

[1] These numbers are remeasured by the length unit $L=219.04$km and velocity unit $V = L/T$, where $T = 5.385$h. A similar number near (216) Kleopatra will be expressed in the same way. For the asteroid (22) Kalliope, $L=191.94$km, $T = 4.1480$h. For the asteroid (433) Eros, $L=34.4$km, $T = 5.27$h.

Symmetrically, when the Jacobian constant is increasing to $1.5149 \times 10^3 \cdot m^2 \cdot s^{-2}$, another cuspidal point arises, and the periodic ratio is 2.0697:1, which is marked as Cuspidal Point 2. After crossing this point, the periodic ratio increases rapidly and converges to 3:1 finally. Here we find the relationship between periodic ratio and Jacobian constant. Between cuspidal point 1 and cuspidal point 2, it is the resonant status of the periodic orbit families. The results implies that the periodic ratios for a fixed resonant status have an infimum and a supremum . The 2:1 resonance in Fig. 2, has the infimum 1.9129:1 and the supremum 2.0697:1.

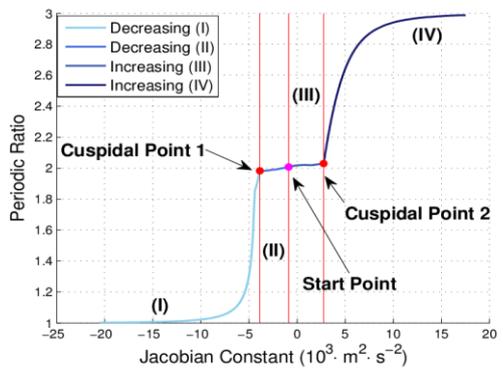

(a) periodic ratio

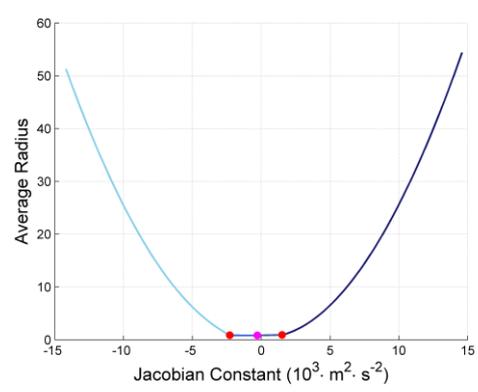

(b) Average Radius

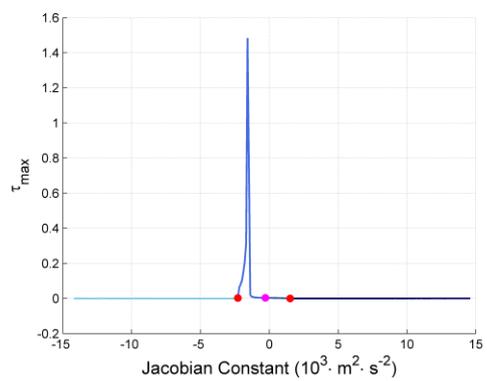

(c) Maximal torsion

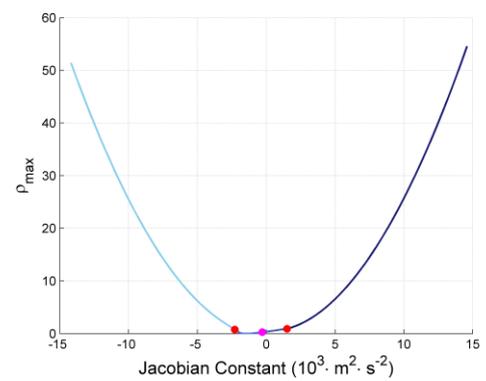

(d) Maximal radius of curvature

**Fig. 2** Four different characteristics of the periodic orbit family near (216) Kleopatra, where the bidirectional continuation is engaged.

Several critical periodic orbits in these four stages are shown in Fig. 3, which provides intuitional and solid evidence to demonstrate the conclusion in Corollary 1. When the Jacobian constant is decreasing, the periodic orbit starts to stretch out (see Fig. 3(b)), and then, it lays down in the equatorial plane and gradually evolves into a circle (see Fig. 3(a)). In contrast, when the Jacobian constant is increasing, the periodic orbit starts to wrap together (see Fig. 3(c)), and then, it converges to a nearly circular orbit with a multiplicity of three (see Fig. 3(d)), wherein this multiplicity cannot be distinguished clearly due to the low pixel resolution.

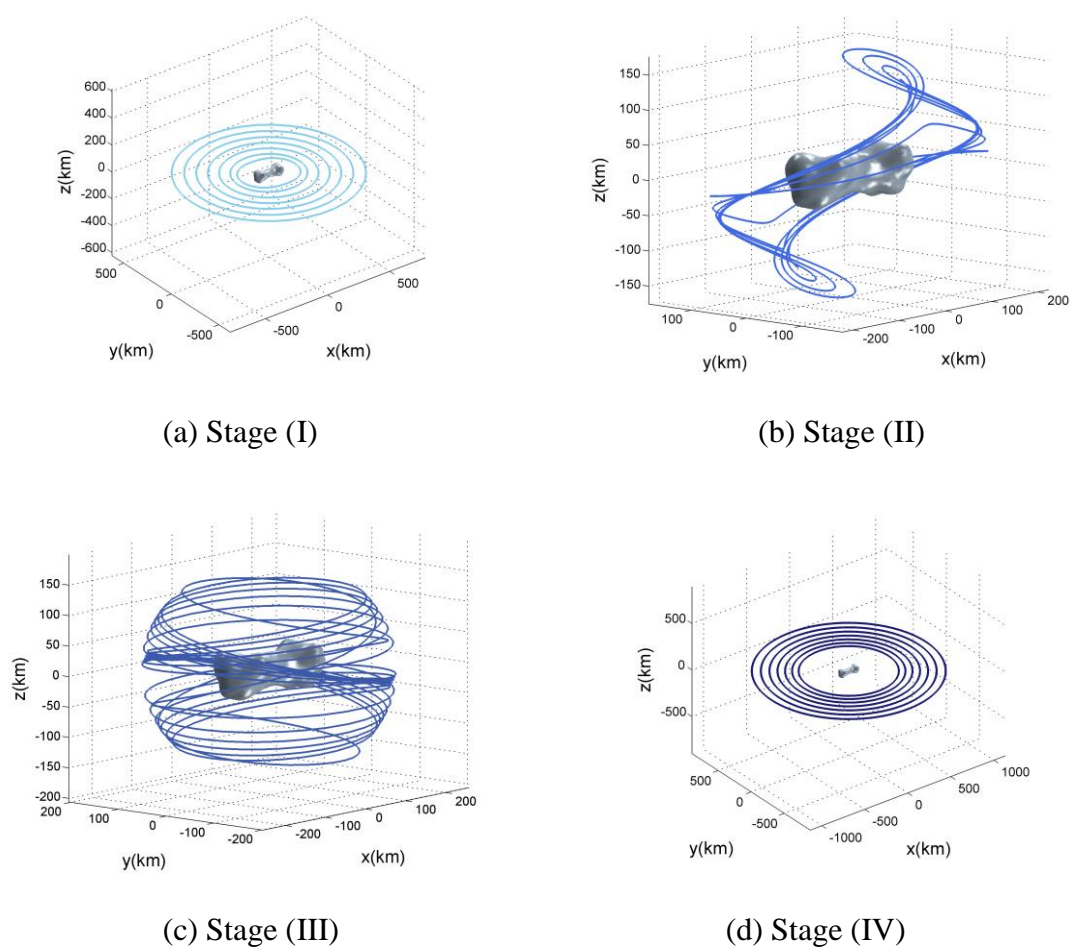

(a) Stage (I)          (b) Stage (II)

(c) Stage (III)        (d) Stage (IV)

**Fig. 3** Representative periodic orbits in four different stages in Figure (2).

Considering the specificity of these two cuspidal points in Fig. 2(a), we analyse the Floquet multipliers of the periodic orbits nearby and find that tangent bifurcation and pseudo tangent bifurcation arise at Cuspidal Points 1 and 2. To be specific, Fig. 4 delineates the transform routine of the Floquet multipliers.

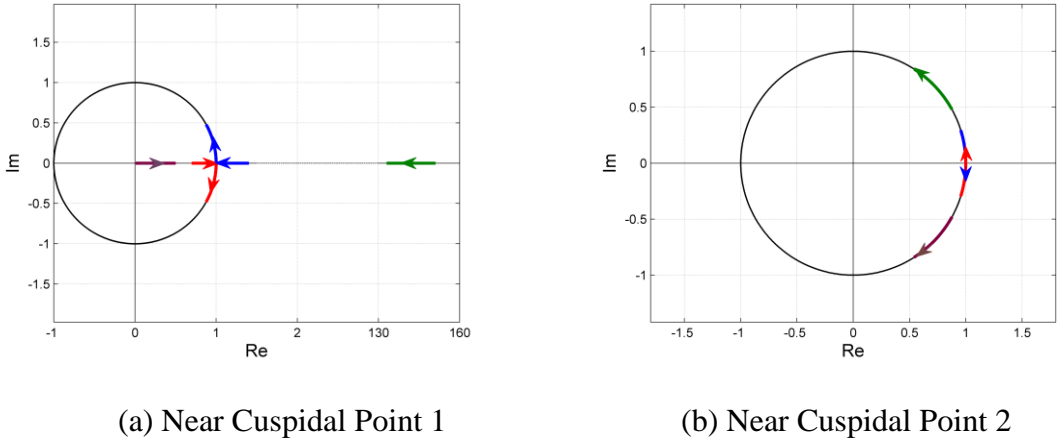

(a) Near Cuspidal Point 1          (b) Near Cuspidal Point 2

**Fig. 4** The transform routine of the Floquet multipliers for the periodic orbits close to Cuspidal Points 1 and 2 in Figure (2).

Recalling the classification in Fig. 1, we can determine that the topological structure transfer order follows the routine of Case P3→Case P6→Case P4 near Cuspidal Point 1 with a decrease in the Jacobian constant, which makes it tangent bifurcation. The periodic orbit at Cuspidal Point 1 exactly belongs to Case P6. Similarly, the topological structure transfer order follows the routine of Case P2→Case P5→Case P2 near Cuspidal Point 2 with an increase in the Jacobian constant, which corresponds to the pseudo tangent bifurcation. The periodic orbit at Cuspidal Point 2 also belongs to Case P5.

**4.1.2 (22) Kalliope**

Close to the asteroid (22) Kalliope, when continuing the periodic orbit with the initial position of [0.0270, 0.8221, −0.2167] and velocity of [5.1072, −1.1096, 3.7380], we find that this orbit will evolve into an almost circular orbit with a change in the Jacobian constant along two directions.

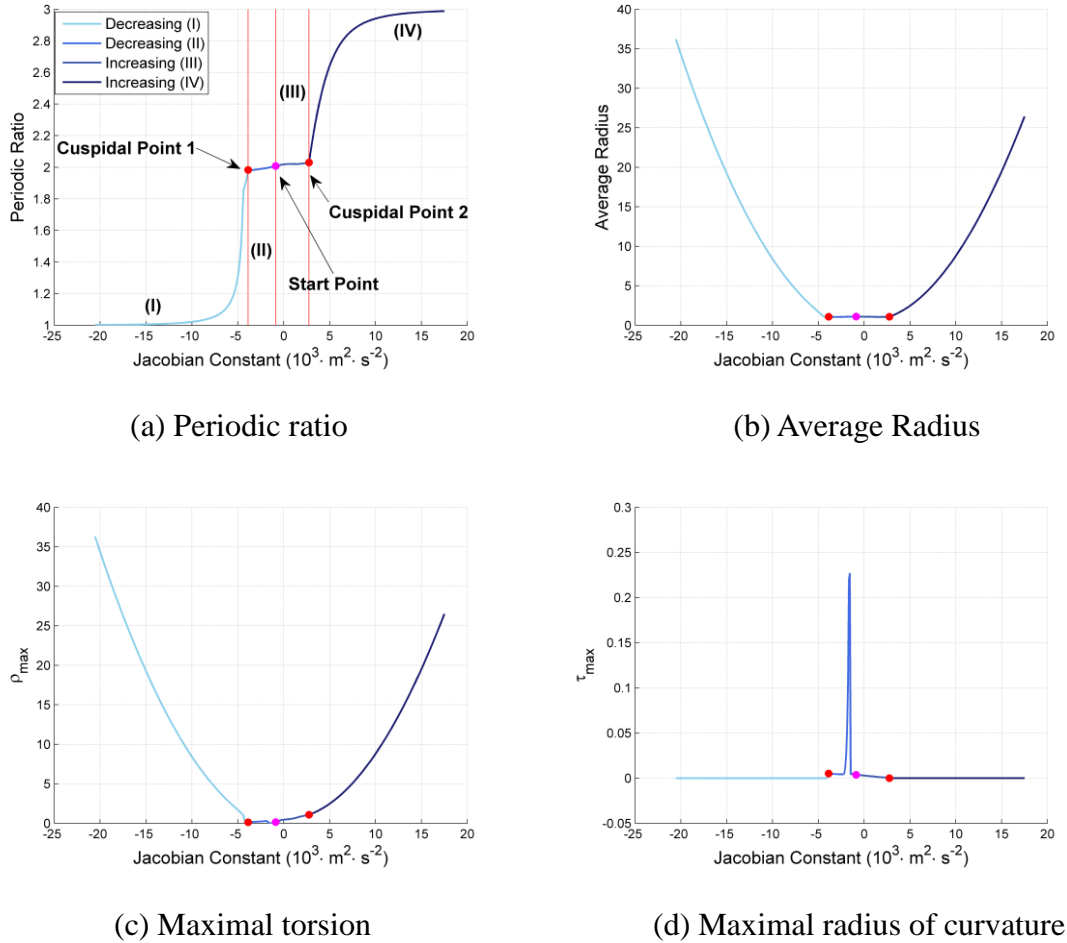

(a) Periodic ratio

(b) Average Radius

(c) Maximal torsion

(d) Maximal radius of curvature

**Fig. 5** Four different characteristics in the periodic orbit family near (22) Kalliope, where the bidirectional continuation is engaged.

Figure 5 shows the four different characteristics in this periodic orbit family: the periodic ratio, average radius, maximal torsion and maximal radius of curvature. Similar to the periodic orbit family mentioned before, a cuspidal point occurs during the decreasing continuation at the position of [−3.8695, 1.9829] in Fig. 5(a), which is marked as Cuspidal Point 1. As the decreasing continuation goes on, the periodic ratio converges to 1:1 gradually. Another cuspidal point shows up during the increasing continuation, which is located at [2.7605, 2.0301] and is marked as Cuspidal Point 2, and the periodic ratio approximates 3:1 finally. We also divide the whole procedure into four stages by these two cuspidal points and the starting point, which are distinguished in the other subgraphs. It is easy to determine that the cuspidal points coincide with the turning points in Fig. 5(b)~5(d).

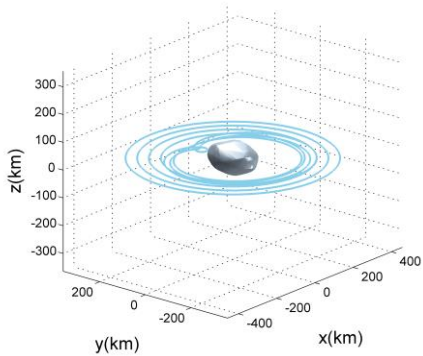

(a) Stage (I)

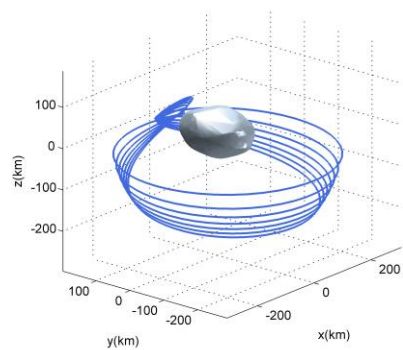

(b) The front part of Stage (II)

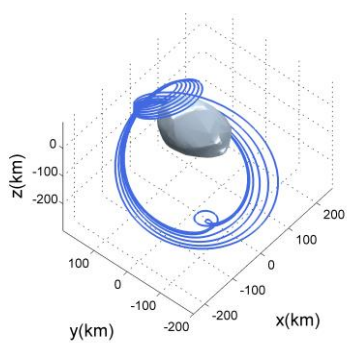

(c) The latter part of Stage (II)

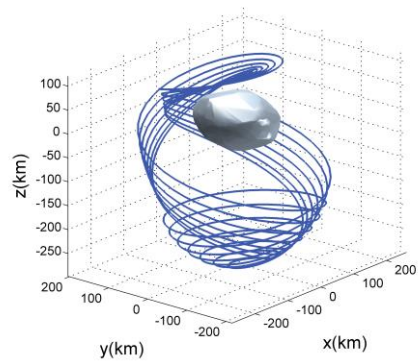

(d) The front part of Stage (III)

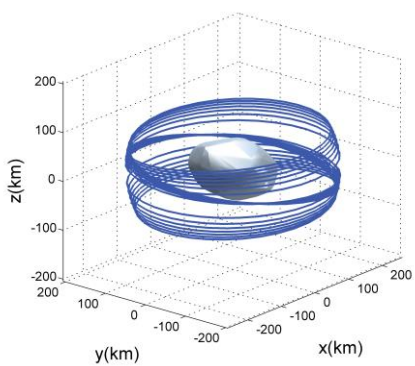

(e) The latter part of Stage (III)

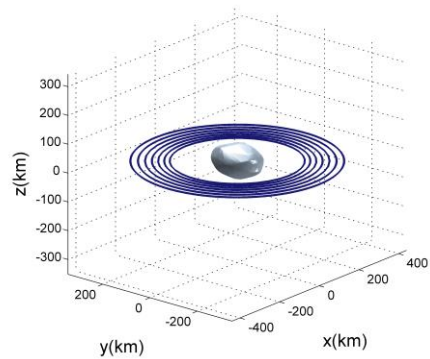

(f) Stage (IV)

**Fig. 6** Representative periodic orbits in four different stages in Figure (5).

Several representative periodic orbits are selected to show this bidirectional continuation, where the front part and the latter part in Stage (II) and Stage (III) are presented separately to show the transformation more explicitly (see Fig. 6(b)~6(e)). Similar to the description in Corollary 1, the periodic orbit evolves into an almost circular orbit when the Jacobian constant is decreasing (see Fig. 6(a)) or increasing (see Fig. 6(f)).

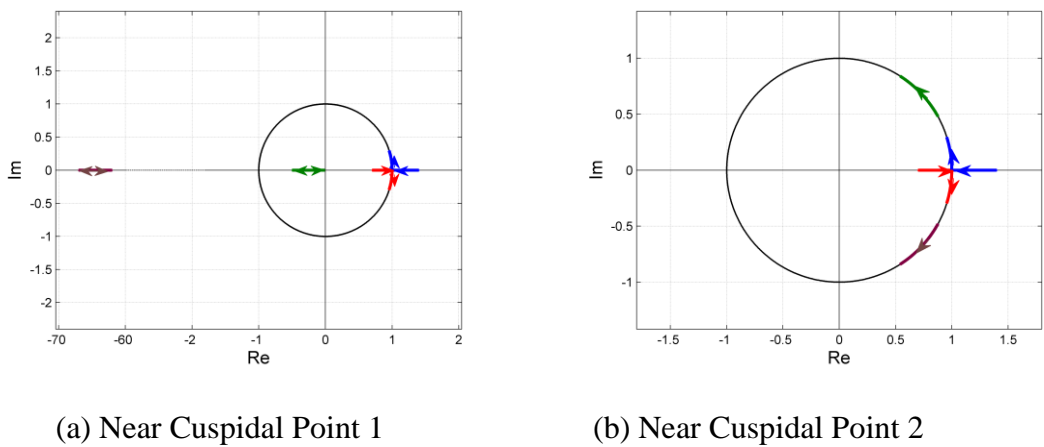

(a) Near Cuspidal Point 1         (b) Near Cuspidal Point 2

**Fig. 7** The transform routine of the Floquet multipliers for the periodic orbits close to Cuspidal Points 1 and 2 in Figure (5).

When analysing the variety of Floquet multipliers and combining the classification in Fig. 1, we can determine that pseudo tangent bifurcation and tangent bifurcation appear at Cuspidal Point 1 and 2, respectively. Fig. 7 demonstrates the variations in the Floquet multipliers.

### 4.1.3 (433) Eros

Near the surface of asteroid (433) Eros, the periodic orbit with the initial position [0.0847, −0.6599, −0.1056] and initial velocity [−5.8643, −0.1782, −1.9257] is found to admit convergence along two directions.

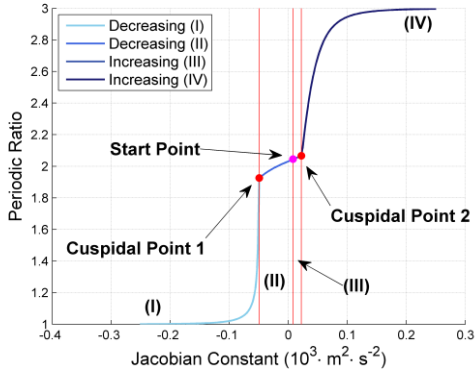
(a) Periodic ratio

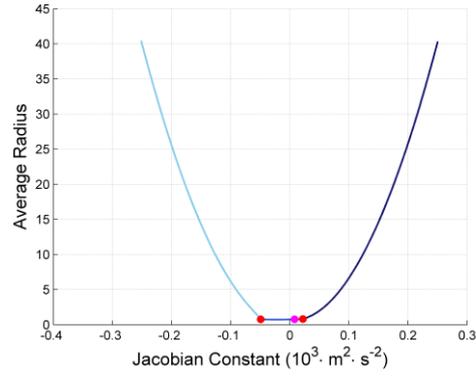
(b) Average Radius

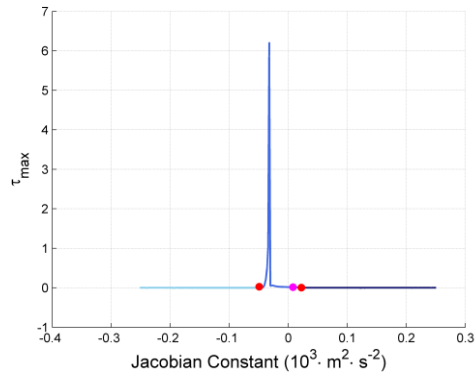
(c) Maximal torsion

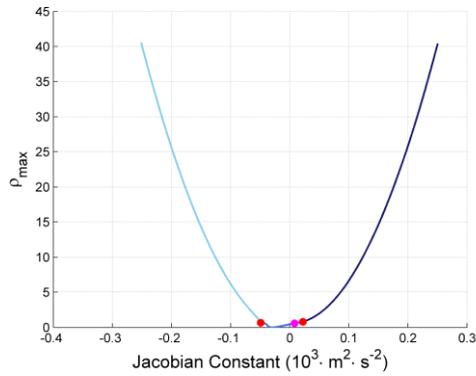
(d) Maximal radius of curvature

**Fig. 8** Four different characteristics of the periodic orbit family near (433) Kleopatra.

After analysing the periodic ratio in Fig. 8(a), two cuspidal points are marked, one in the decreasing direction and the other in the increasing direction. With the help of these two critical points and the starting point, we can divide this periodic orbit family into four parts, which correspond to the four stages in Fig. 8(a). Furthermore, the variations in the average radius, maximal torsion and maximal radius of curvature versus the Jacobian constant in this periodic orbit family are also plotted in Fig. 8(b)~8(d). Clearly, the cuspidal points in Fig. 8(a) correspond to the turning points in the other subgraphs.

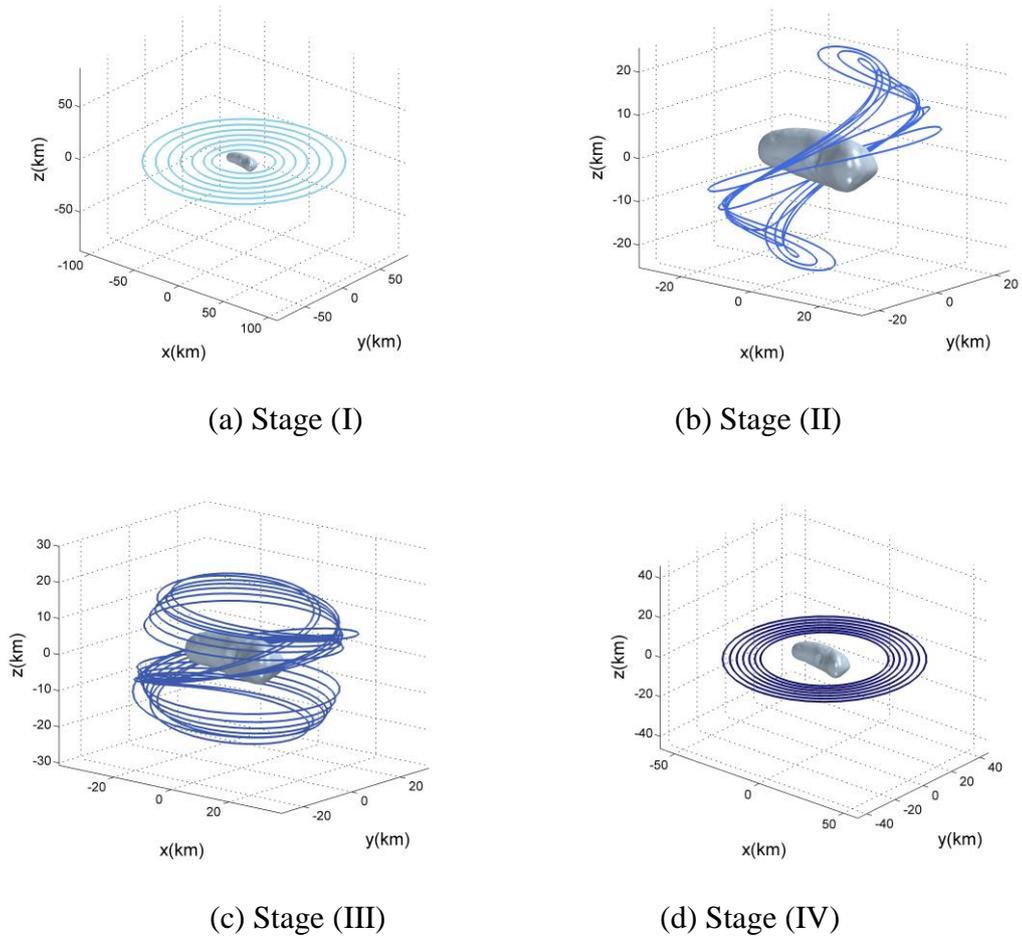

(a) Stage (I)　　　　　　　　　　(b) Stage (II)

(c) Stage (III)　　　　　　　　　(d) Stage (IV)

**Fig. 9** Representative periodic orbits in four different stages in Figure (8).

Via numerical simulation, we plot several periodic orbits of the four stages in Fig. 9, where the transformation of the periodic orbit during the continuation can be exhibited more intuitively. Moreover, the changes in the Floquet multipliers for the periodic orbits near the cuspidal points are shown in Fig. 10, which implies that the pseudo tangent bifurcations occur at these two cuspidal points.

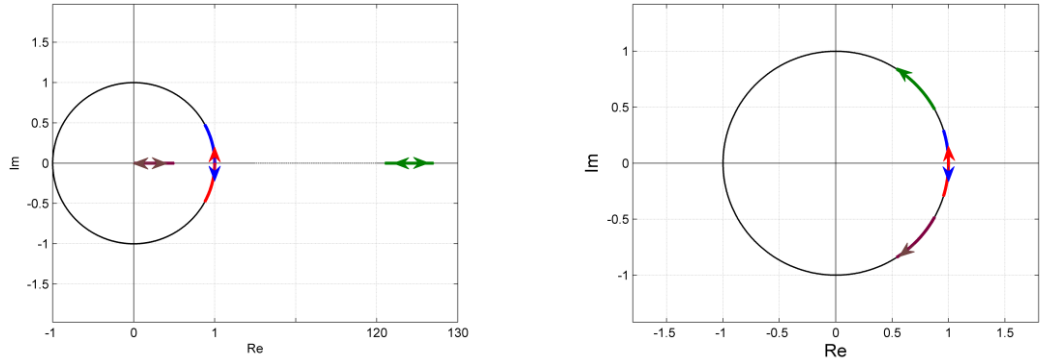

**Fig. 10** The topological structure transform routine close to Cuspidal Points 1 and 2 in Figure (8).

### 4.2 Increasing-directional continuation

In this case, a periodic orbit will converge to a nearly circular orbit when the Jacobian constant is increasing, but it will vanish into a point gradually when the Jacobian constant is decreasing. With further calculation, we determine that this point is exactly one of the equilibrium points of the asteroid.

#### 4.2.1 (216) Kleopatra

Among the many periodic orbit families near (216) Kleopatra, we found that one of them, with initial position of [−0.0758, −0.6003, −0.0281] and velocity of [−5.6601, 0.7846, −2:5506], belongs to this case. Additionally, the mentioned equilibrium point that is converged into by this periodic orbit is unstable and belongs to Case 5 according to the classification in Table 1.

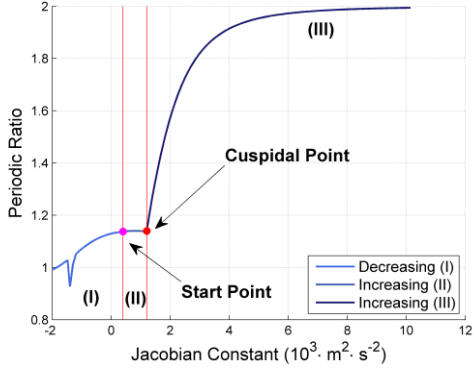 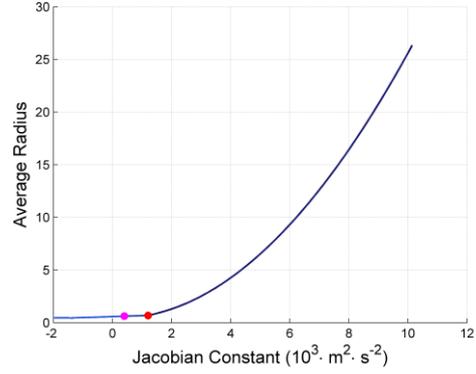

(a) Periodic ratio　　　　　　　　　(b) Average Radius

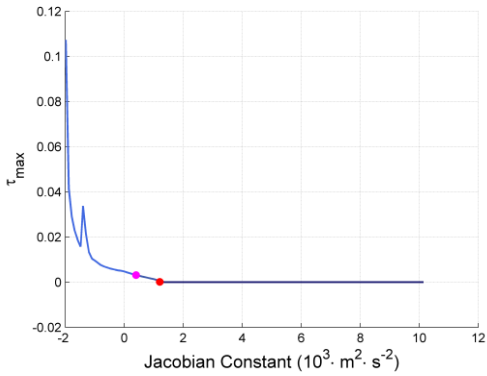 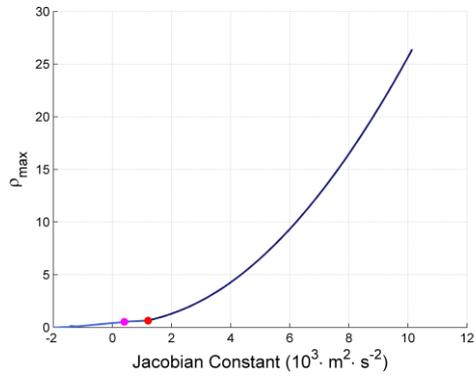

(c) Maximal torsion　　　　　　　　(d) Maximal radius of curvature

**Fig. 11** Four different characteristics in the periodic orbit family near (216) Kleopatra, where the increasing-directional continuation is engaged.

Figure 11 shows the variation in the periodic ratio, the average of the radius, the maximal radius of curvature and the maximal torsion, according to the approach in Fig. 2. It is easy to determine that when the Jacobian constant is increasing and reaches $1.2027 \times 10^3 \cdot m^2 \cdot s^{-2}$, a cuspidal point shows up, and the periodic ratio is 1.14:1 there, which is marked in Fig. 11(a). Afterward, the increase in the periodic ratio accelerates suddenly and converges to 2:1 finally. However, along the other decreasing direction, the periodic orbit will vanish into an equilibrium point when the Jacobian constant reaches the value of $-1.9772 \times 10^3 \cdot m^2 \cdot s^{-2}$, and the periodic ratio stops at 0.9918:1. Combining the starting point and the cuspidal point, we can divide the whole procedure into three stages (see Fig. 11(a)).

They are also distinguished in Fig. 11(b)∼11(d), which presents the change tendency of the average radius, the maximal torsion and the maximal radius of curvature, respectively. Coincidentally, the cuspidal point matches the turning points in these figures. Similar to the conclusion in Section 3, with the increase in the Jacobian constant and after crossing the cuspidal points, the average radius and the maximal radius of curvature increase steadily, and the maximal torsion decreases to zero.

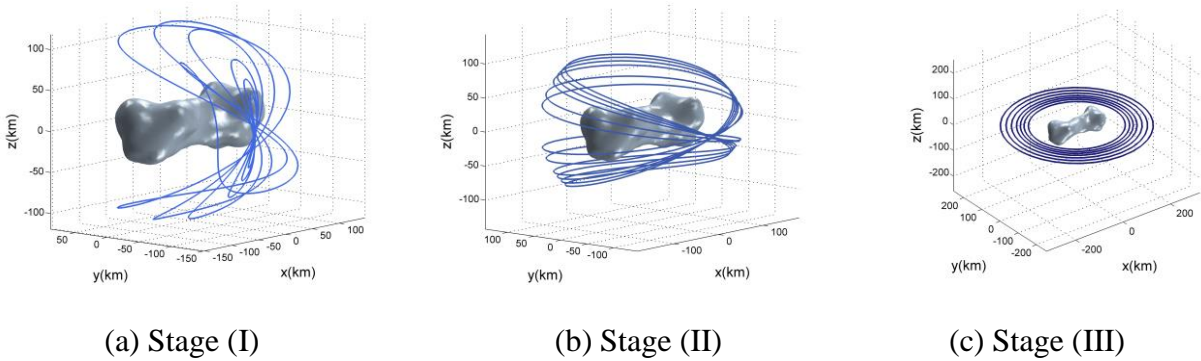

(a) Stage (I)  (b) Stage (II)  (c) Stage (III)

**Fig. 12** Representative periodic orbits of the three different stages in Figure (11).

Figure 12 shows several representative periodic orbits in three stages. When the Jacobian constant increases, the periodic orbit starts to wrap together (see Fig. 12(b)), and then, it converges to a nearly circular orbit with a multiplicity of two (see Fig. 3(c)). On the other hand, when the Jacobian constant decreases, the periodic orbit shrinks to a smaller orbit until it vanishes to a point, which is exactly an unstable equilibrium point of (216) Kleopatra and belongs to Case 5 according to the topological manifold classifications in Table 1.

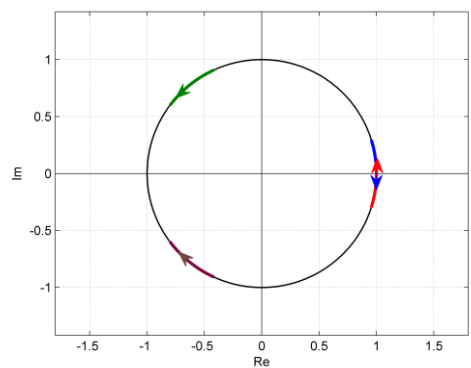

**Fig. 13** The topological structure transform routine close to the cuspidal point in Figure (11).

Moreover, Floquet multipliers for the periodic orbits near the cuspidal point are calculated, and their variations are plotted in Fig. 13. It is easy to determine that the pseudo tangent bifurcation arises here, and the topological structure follows the routine of Case P2→Case P5→Case P2. The critical state of Case P5 exactly corresponds to the periodic orbit at the cuspidal point.

### 4.2.2 (22) Kalliope

For the asteroid (22) Kalliope, we employ the periodic orbit seed with an initial position of [0.6771, 0.1570, 0.1329] and velocity of [2.0190, -7.3259, -2.3770] to expand our discussion.

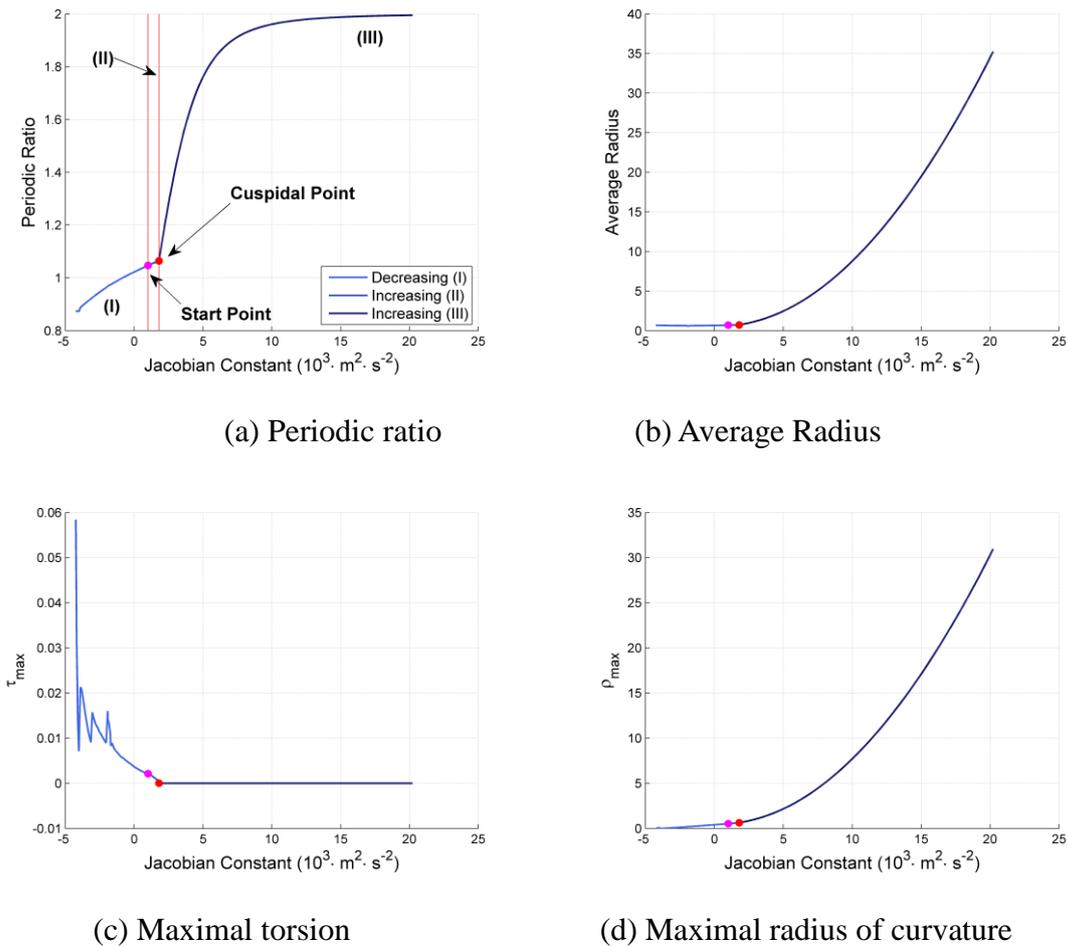

(a) Periodic ratio         (b) Average Radius

(c) Maximal torsion        (d) Maximal radius of curvature

**Fig. 14** Four different characteristics in the periodic orbit family near (22) Kalliope, where the increasing-directional continuation is engaged.

Similar to the previous method, we analyse the change tendency of four characteristics in this periodic orbit family. One cuspidal point is scattered in Fig. 14(a), and it overlaps the turning points in Fig. 4(b)~14(d). Clearly, while the Jacobian constant is increasing, the periodic ratio converges to 2:1, the average radius and the maximal radius of curvature enlarges steadily, and the maximal torsion falls to zero. With the help of the cuspidal point and the starting point, the three stages are naturally distinguished.

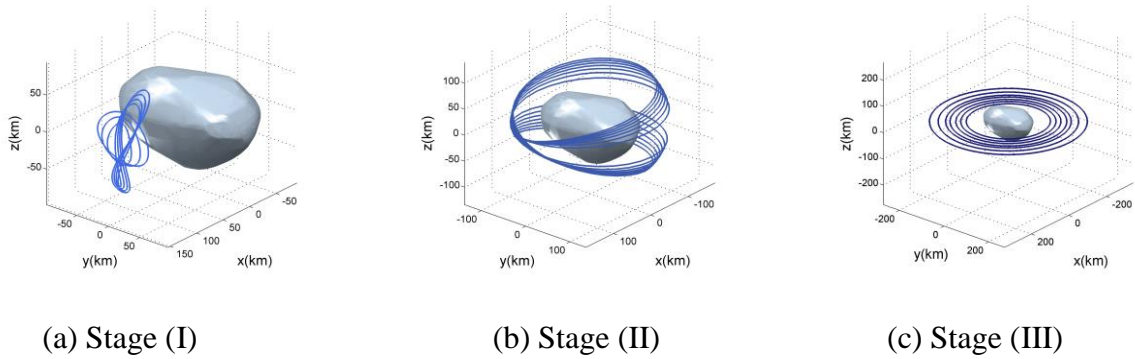

(a) Stage (I)　　　　　　　(b) Stage (II)　　　　　　　(c) Stage (III)

**Fig. 15** Representative periodic orbits in three different stages in Figure (14).

Some periodic orbits are selected to exhibit the three stages in Fig. 15, which can help to clearly understand the convergence to a circular orbit in the increasing direction and the shrinking to an unstable equilibrium point of Case 5 (see Table 1) in the decreasing direction. After calculating the Floquet multipliers near the cuspidal point, tangent bifurcation is discovered, and the transfer routine is plotted in Fig. 16.

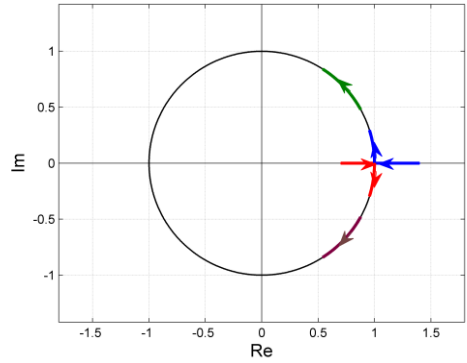

**Fig. 16** The transform routine of the Floquet multipliers for the periodic orbits close to the cuspidal point in Figure (14).

### 4.2.3 (433) Eros

Near the asteroid (433) Eros, the periodic orbit with the initial state [−0.0083, 0.5025, 0.0106] and initial velocity [4.7371, 0.0890, −2.3103] can evolve into being close to a circle with an increase in the Jacobian constant but shrink to a point with a decrease in the Jacobian constant, as in the earlier example.

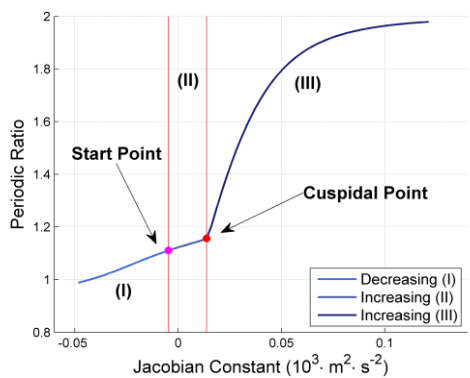
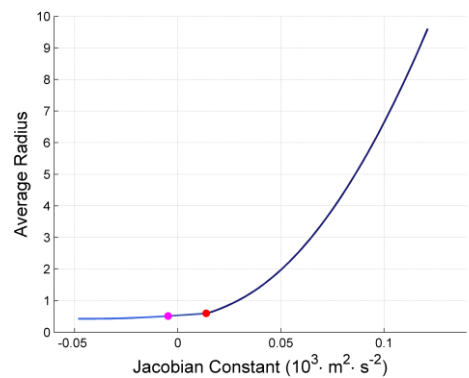

(a) Periodic ratio  (b) Average radius

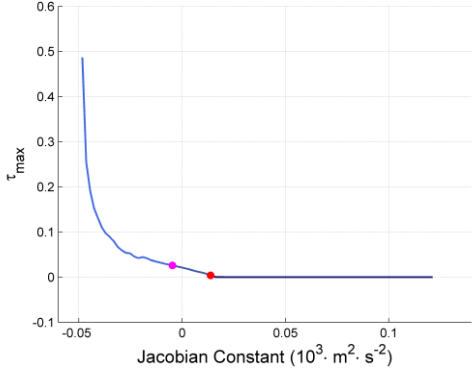
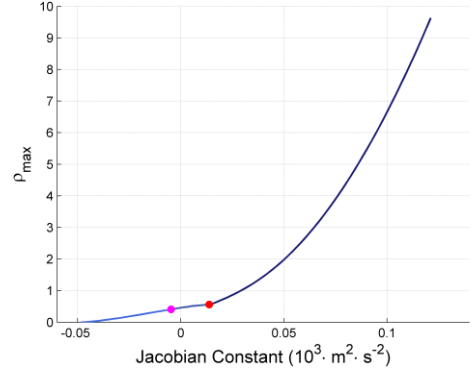

(c) Maximal torsion

(d) Maximal radius of curvature

**Fig. 17** Four different characteristics of the periodic orbit family near (433) Kleopatra.

Figure 17 shows four different characteristics of this periodic orbit family. In Fig. 17(a), a cuspidal point emerges when the Jacobian constant is $0.0138 \times 10^3 m^2 s^{-2}$ and the periodic ratio is 1.1566. The periodic orbit that corresponds to this point also bring about the turning points for the other three characteristics in Fig. 17(b)~17(d). When combining the starting point, we can divide the whole procedure into three stages, where several periodic orbits are plotted in Fig. 18 to show the convergence in a visual way.

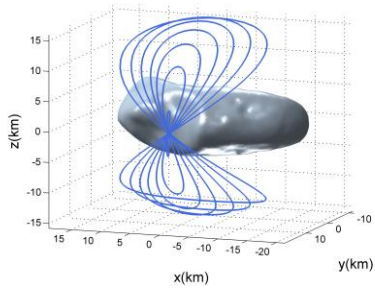
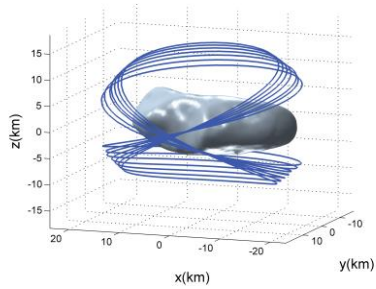
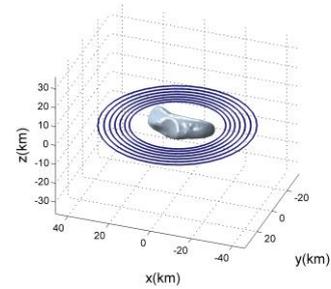

(a) Stage (I)          (b) Stage (II)          (c) Stage (III)

**Fig. 18** Representative periodic orbits in four different stages in Figure (17).

At the same time, the variations in the Floquet multipliers for the periodic orbits near the cuspidal point are analysed in Fig. 19, and the pseudo tangent bifurcation is found to be exactly at the cuspidal point.

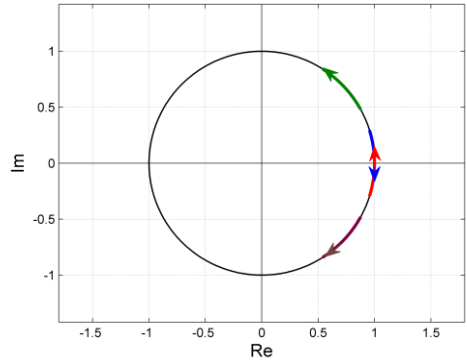

**Fig. 19** The topological structure transform routine close to the cuspidal point in Figure (17).

## 4.3 Decreasing-directional continuation

In the third case, the periodic orbit will converge to an almost circular orbit with a decrease in the Jacobian constant. However, when the Jacobian constant increases, it will become stuck at the local extremum.

### 4.3.1 (216) Kleopatra

We take the periodic orbit family with the initial position of [0.0376, −0.8485, 0.0287] and the initial velocity of [−6.3246, −0.3505, −2.5520] as an example.

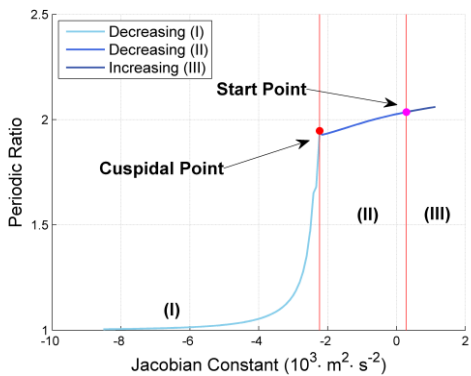  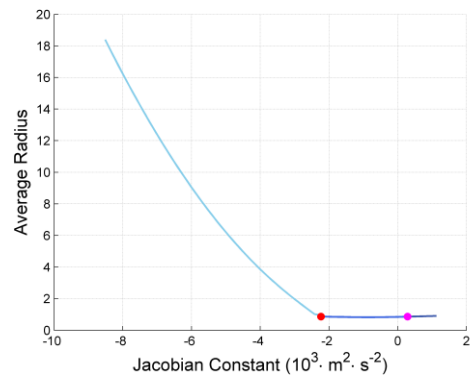

(a) Periodic ratio  (b) Average radius

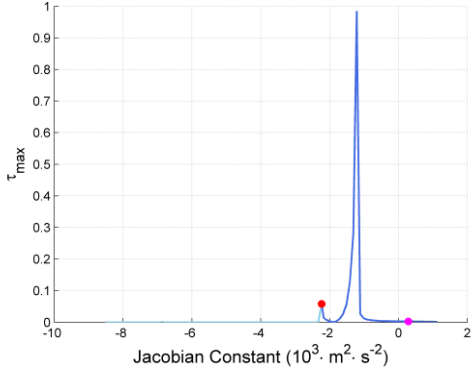
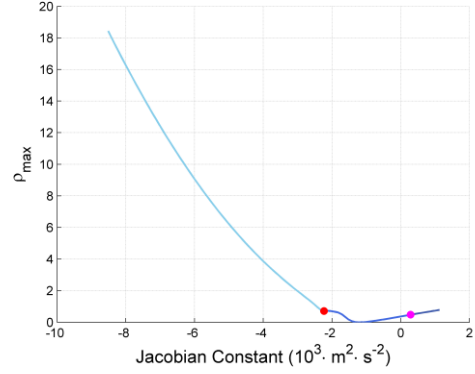

(c) Maximal torsion  (d) Maximal radius of curvature

**Fig. 20** Four different characteristics in the periodic orbit family near (216) Kleopatra, where the decreasing-directional continuation is engaged.

The variations in the four characteristics for this periodic orbit family are shown in Fig. 20. When the Jacobian constant decreases and reaches $-2.2315 \times 10^3 \cdot m^2 \cdot s^{-2}$, a cuspidal point occurs (see Fig. 20(a)), and the periodic ratio is 1.9465:1. After crossing this point, the periodic ratio decreases rapidly and converges to 1:1 finally. In the other direction of continuation, when the Jacobian constant increases and reaches $1.129 \times 10^3 \cdot m^2 \cdot s^{-2}$, the continuation stops, and the periodic ratio also remains at the value of 2.0601:1. The three stages of the whole procedure, which are divided by the cuspidal point and starting point, are marked. It is noteworthy that the cuspidal point represents the turning points for the average radius (see Fig. 20(b)), the maximal torsion (see Fig. 20(c)) and the maximal radius of curvature (see Fig. 20(d)).

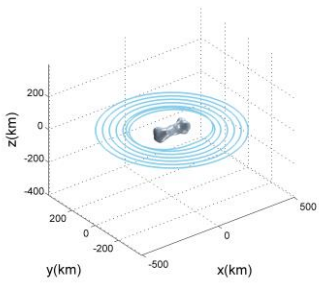
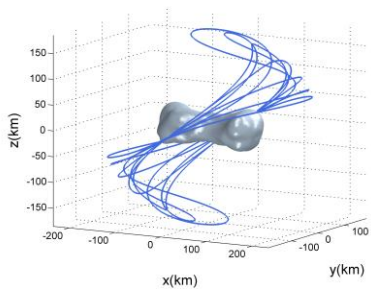
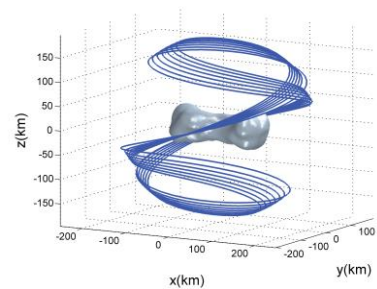

(a) Stage (I)  (b) Stage (II)  (c) Stage (III)

**Fig. 21** Representative periodic orbits in three different stages in Figure (20).

Some representative periodic orbits in these three stages are plotted in Fig. 21. It is not difficult to determine that as the Jacobian constant decreases, the periodic orbit will stretch out (see Fig. 21(b)), and then, it lays down in the equatorial plane and converges to a circle gradually (see Fig. 21(a)). However, when the Jacobian constant increases, the periodic will stop at a certain shape and cannot evolve more.

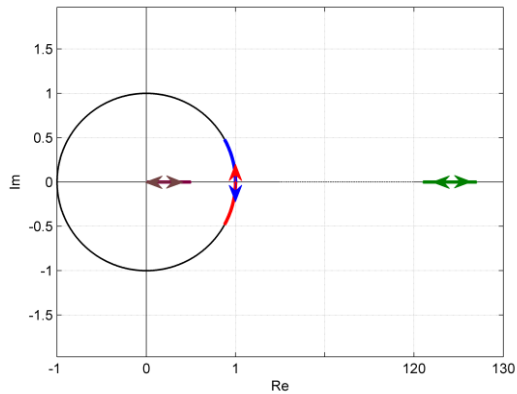

**Fig. 22** The topological structure transform routine close to the cuspidal point in Figure (20).

After calculating the Floquet multipliers of the periodic orbits near the cuspidal points, we can recognise that pseudo tangent bifurcation exists there. The variations in the Floquet multipliers are described in Fig. 22. The topological structure follows the routine of Case P4→Case P6→Case P4, and the periodic orbit at the cuspidal point exactly corresponds to the Case P6.

### 4.3.2 (22) Kalliope

The convergence in Corollary 1 is also detected in the period orbit family with the initial position of [−0.9890, −0.0428, 0.0497] and velocity of [−0.5033, 7.1442, −3.2478] near (22) Kalliope when the Jacobian constant is decreasing. However, as the Jacobian constant increases, the continuation will stop at the value of $1.0810 \times 10^3 \cdot m^2 \cdot s^{-2}$.

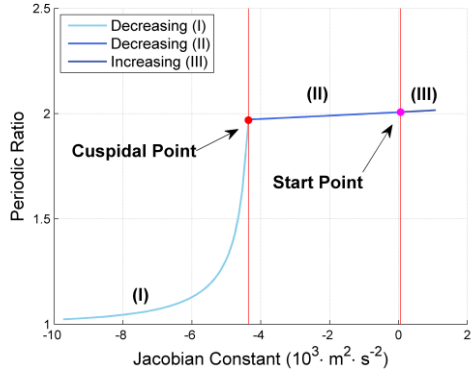 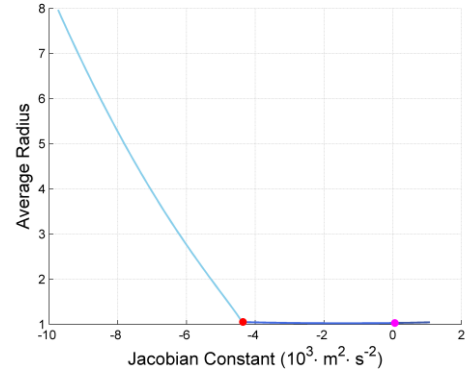

(a) Periodic ratio    (b) Average radius

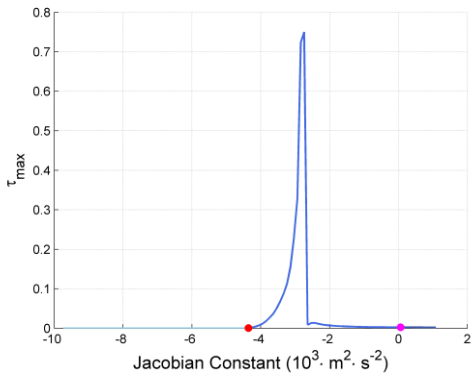 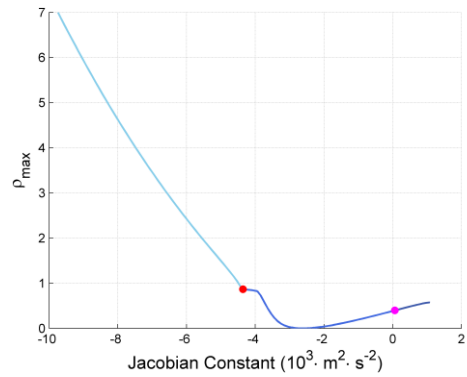

(c) Maximal torsion    (d) Maximal radius of curvature

**Fig. 23** Four different characteristics in the periodic orbit family near (22) Kalliope, where the bidirectional continuation is engaged.

Figure 23 shows the change tendency of four characteristics in this periodic orbit family. When the Jacobian constant decreases, the periodic ratio decreases and comes to a cuspidal point with the value of 1.9692, where the Jacobian constant is $-4.3590 \times 10^3 \cdot m^2 \cdot s^{-2}$. Additionally, this cuspidal point coincides with the turning points in other subgraphs. Fig. 23(a) introduces three stages, which are divided by the cuspidal point and the starting point, and several periodic orbits are selected to show the transformation (see Fig. 24) during the continuation. In these figures, it is explicitly shown that the periodic orbit converges to a nearly circular orbit as the Jacobian constant decreases. In the other direction of the continuation, it becomes stuck, which is also shown in Fig. 24(c).

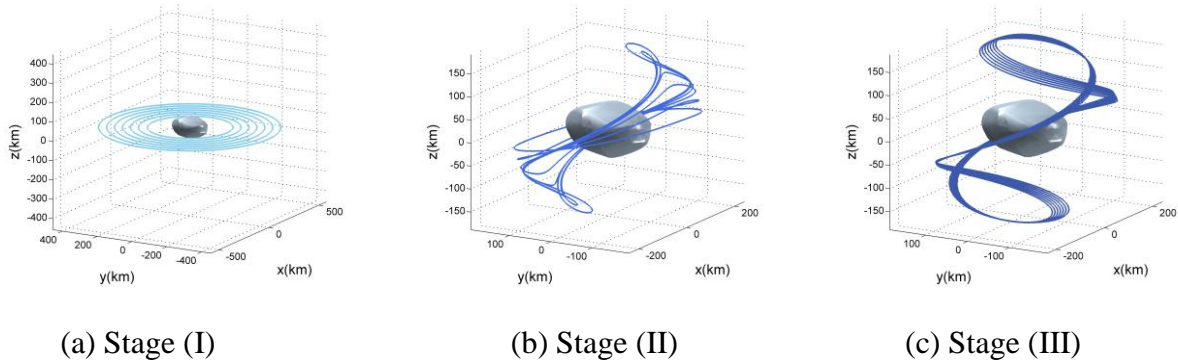

(a) Stage (I)          (b) Stage (II)          (c) Stage (III)

**Fig. 24** Representative periodic orbits in three different stages in Figure (23).

The same as the earlier methods, the Floquet multipliers of the periodic orbits near the cuspidal points are analysed, and their variations are sketched in Fig. 25. Combining the result in Jiang et al. (2015b), we can determine that tangent bifurcation occurs here. Additionally, the topological structure follows the routine of Case P3→Case P6→Case P4.

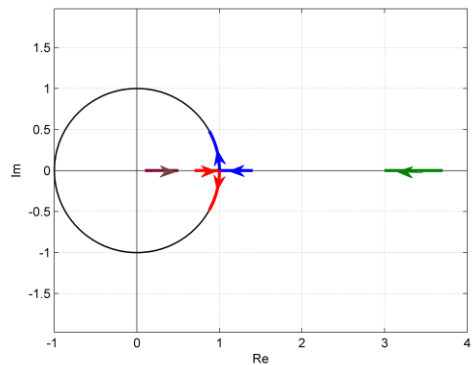

**Fig. 25** The transform routine of the Floquet multipliers for the periodic orbits close to the cuspidal point in Figure (23).

**4.3.3 (433) Eros**

For the asteroid (433) Eros, we chose the periodic orbit with the initial position [0.1477, −0.7107, −0.0528] and initial velocity [−4.8432, −1.1420, 2.3570] to conduct the discussion.

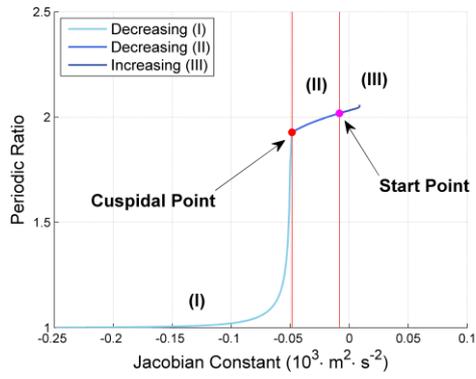
(a) Periodic ratio

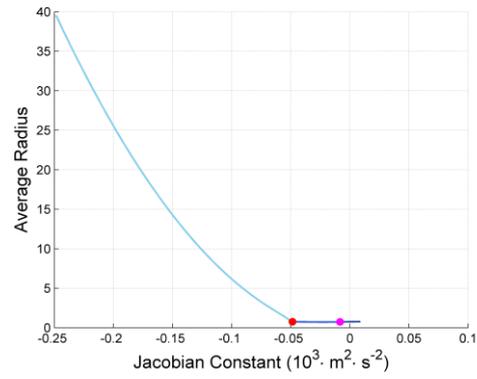
(b) Average radius

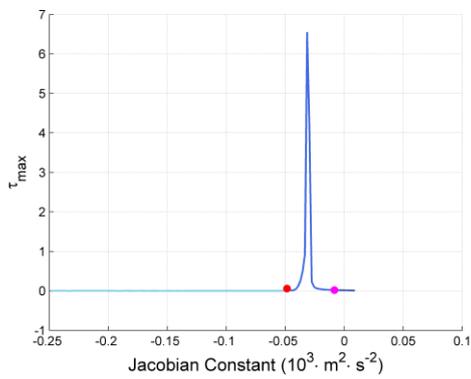
(c) Maximal torsion

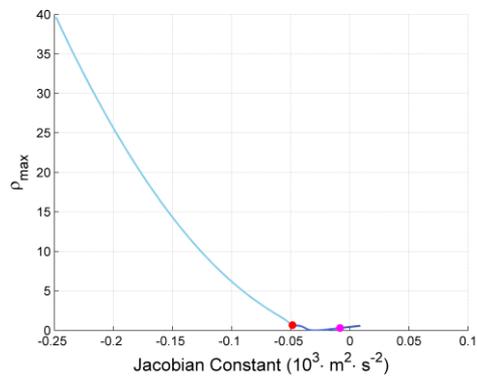
(d) Maximal radius of curvature

**Fig. 26** Six different characteristics in the periodic orbit family near (433) Kleopatra.

In Fig. 26, four different characteristics of this periodic orbit family are plotted, where a cuspidal point, starting point and three stages are marked in Fig. 26(a). Clearly, the cuspidal point also represents the turning point in the other subgraphs. Several representative periodic orbits in three different stages in Fig. 26(a) are shown in Fig. 27, which can apparently show the variations in the periodic orbit during this continuation.

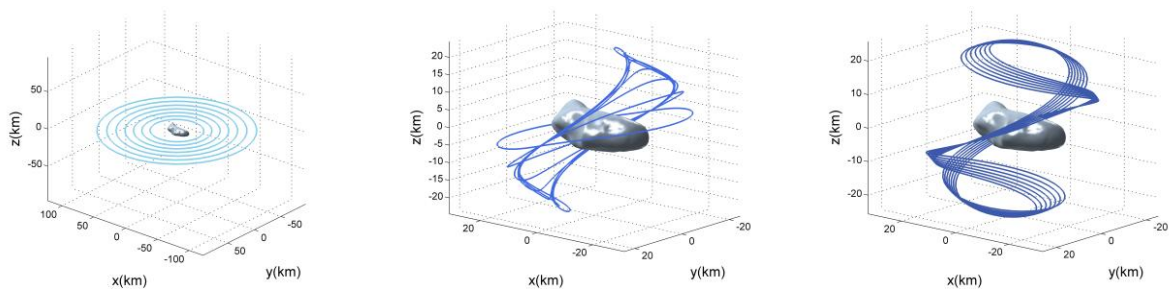

| (a) Stage (I) | (b) Stage (II) | (c) Stage (III) |

**Fig. 27** Representative periodic orbits in three different stages in Figure (26).

Moreover, at the cuspidal point, the topological structure also changes, which can be confirmed from Fig. 28. According to the classification in Fig. 1, we can distinguish it as pseudo tangent bifurcation, and the topological structure follows the routine of Case P→Case P5→Case P2.

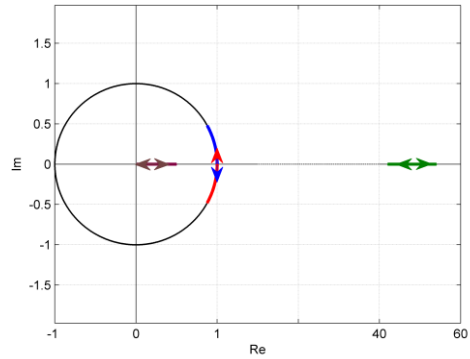

**Fig. 28** The topological structure transform routine close to the cuspidal point in Fig. 26.

### 4.4 Comparison of the periodic orbit families

Although there are many dissimilarities in these numerical simulations among the different periodic orbit families near different asteroids, such as the dispersion of the periodic ratios and the distinctions in the stabilities, we can still find several commonalities.

First, the periodic ratio will present sharp changing during the continuation of the Jacobian constant, where the cuspidal points form and the (pseudo) tangent bifurcation arises. In addition, these cuspidal points in the periodic ratio coincide with the turning points of the average radius, the maximal torsion and the maximal radius of curvature. Furthermore, the overall trend in the periodic ratio agrees with the one of the Jacobian constants. This finding means that the periodic ratio increases (or decreases) with an increase (or decrease) in the Jacobian constant overall. Finally, if a periodic orbit shrinks to a point during the continuation, it could be an unstable equilibrium point of the corresponding asteroid and belongs to Case 5 in Table 1.

## 5. Conclusions

In this paper, we first discover the convergence for a periodic orbit family theoretically and numerically when the Jacobian constant changes over a wide range. Through theoretical derivation, we found that if the continuation of a periodic orbit based on the Jacobian constant could always be conducted and will not fall into a valley of a local extremum or the periodic orbit will not vanish into a point, this periodic orbit will converge to an almost circular periodic orbit with different multiplicities in the equatorial plane. Moreover, through numerical calculation, we utilise the radius of curvature and the torsion to quantise the characteristics of a periodic orbit, which helps us understand the variation in a single periodic orbit family during continuation from the geometric aspect and provides a brand-new analysis perspective.

As an application of the result, three asteroids, (216) Kleopatra, (22) Kalliope and (433) Eros, are studied, and some representative period orbit families nearby are discussed in detail, for which we present the convergence in the bidirectional, increasing-direction and decreasing-direction continuation. Finally, four commonalities among these numerical examples are concluded, which could provide several ideas and proofs for future discussion.

**Acknowledgements**

This research was supported by the National Natural Science Foundation of China (Grant No. 11772356), China Postdoctoral Science Foundation funded project (Grant No. 2017M610875, No. 2018T110092), and China Scholarship Council (visiting scholar, No. 201703170036) .